\documentclass{cta-author}
\usepackage{hyperref}
\usepackage{subfig}
\usepackage{graphicx}
\usepackage{algpseudocode}
{}
{}
{}

\begin{document}

\supertitle{Submission Template for IET Research Journal Papers}

\title{Intrusion Detection Using Mouse Dynamics}

\author{\au{Margit Antal$^{1\corr}$}, \au{El\H{o}d Egyed-Zsigmond $^{2}$}}

\address{\add{1}{Faculty of Technical and Human Sciences, Sapientia University, Calea Sighisoarei 1C, Tirgu Mures, Romania}
\add{2}{Université de Lyon, LIRIS UMR 5205 CNRS, Villeurbanne, France}
\email{manyi@ms.sapientia.ro}}

\begin{abstract}
Compared to other behavioural biometrics, mouse dynamics is a less explored area. General purpose data sets containing unrestricted mouse usage data are usually not available. The Balabit data set was released in 2016 for a data science competition, which against the few subjects, can be considered the first adequate publicly available one.  This paper presents a performance evaluation study on this data set for impostor detection. The existence of very short test sessions makes this data set challenging. Raw data were segmented into mouse move, point and click and drag and drop types of mouse actions, then several features were extracted. In contrast to keystroke dynamics, mouse data is not sensitive, therefore it is possible to collect negative mouse dynamics data and to use two-class classifiers for impostor detection. Both action- and set of actions-based evaluations were performed. Set of actions-based evaluation achieves 0.92 AUC on the test part of the data set. However, the same type of evaluation conducted on the training part of the data set resulted in maximal AUC (1) using only 13 actions. Drag and drop mouse actions proved to be the best actions for impostor detection.
\end{abstract}

\maketitle

\section{Introduction}\label{intro}

Protecting accounts from unwanted access is becoming increasingly important. Online banking, online shopping, or online examinations are part of our everyday routine. Several types of biometrics, such as fingerprint, iris, facial, voice have been already used for this purpose. Unfortunately, they are, on the one hand, intrusive and on the other hand they do not provide continuous protection. Behavioural biometric methods are not intrusive and provide continuous control. 


Several behavioural biometrics have already been studied for intrusion detection. Computer keyboard usage is usually referenced as keystroke dynamics. The way a user types his own password can be used for static (entry point) authentication. These type of systems usually work with a high error rate (5.32\%\cite{MORALES2016}), therefore it is accepted only as an additional biometrics combined with more secure biometrics. Dynamic (continuous) authentication requires modelling the user's typing style. Both static and dynamic authentications require logging the users' typing, which may contain sensitive data. 

Mouse usage is usually referenced as mouse dynamics. In contrast to keystroke dynamics, the study of users' mouse usage does not require sensitive data from the users. Moreover, in web-based environments mouse usage is more frequent than keyboard usage \cite{KRATKY2018}.  This type of behavioural biometrics has already been used for static and dynamic authentications. One of the main limitations
 of some previous studies is that their results are not reproducible due to the unavailability of the data sets. Public data sets would allow us to compare different methods for a given problem in an objective way. This is particularly important in biometric research. Although Ahmed and Traore \cite{AHMED2007} published their data sets - because raw data is not included - their data sets are only suitable for performing certain types of measurements. Despite the fact that it has been written in several papers \cite{SHEN2012}, \cite{SHEN2013}, \cite{SHEN2014} that the result of their research is a public data set for mouse dynamics, in reality such a data set did not exist until 2016, when a data set appeared thanks to the Budapest office of the Balabit company\footnote{https://github.com/balabit/Mouse-Dynamics-Challenge}. This data set includes mouse pointer timing and positioning information; its purpose is to be a benchmark data set in the field of Information Technology security. This data set was first made public during a data science competition\footnote{https://datapallet.io/competition/balabit/detail/}. Unfortunately, the results of the competition are not publicly available.

Another limitation of some previous studies is that experiments were conducted on data collected from a single application, usually specially designed for data collection \cite{GAMBOA2004}, \cite{SHEN2013}, \cite{SHEN2014}. In these applications users were required to perform with the mouse some very specific tasks, therefore these are not natural mouse usages. 

The major contribution of our work are the followings:
\begin{itemize}
\item We present a statistical analysis of the Balabit data set, which contains mouse data collected through general computer usage.
\item We compare action- and set of actions-based evaluations in order to determine the number of mouse actions required for high precision intrusion detection.
\item We present measurements in order to determine which type of mouse action is the most user specific.
\item We took special care of the reproducibility of our measurements and described our experiments with a great attention towards  this objective.

\end{itemize}

Our software and other resources of this research are available at \href{http://www.ms.sapientia.ro/~manyi/mousedynamics/}{http://www.ms.sapientia.ro/\textasciitilde manyi/mousedynamics/}. The rest of this paper is organized as follows. A literature review on mouse dynamics is presented in Section 2. This is followed by the description of the Balabit data set and its statistical analysis. Section 4 presents the methods used in this study, namely our feature extraction and the machine learning algorithm. Experimental results are presented in Section 5. This is followed by discussion and conclusions.

\section{Background and Related Work}

Behavioural biometrics is attracting increasing attention in security research. Keystroke, touch and mouse dynamics belong to this type of biometric. Keystroke dynamics is already used in on-line courses as a method for continuous authentication. However, keystroke data may contain sensitive data such as passwords and user names. In contrast, mouse dynamics data do not contain such sensitive data. Therefore, it is easy to collect and use this kind of information for intruder detection, for example in on-line banking systems.

In mouse dynamics the time taken to detect an intruder is dependent on the number of mouse events, and also on the type of mouse events (mouse click, mouse move, and drag-and-drop).

In this section we review the most important works according to the following aspects:
\begin{itemize}
\item the type of data collection, 
\item data availability and research reproducibility,
\item the modelling unit the features are extracted from.
\end{itemize}


A distinction must be made regarding types of data collection. Specific data is collected when we observe the usage of a certain application (web or desktop), while general data is generated when a user performs daily tasks on their operating system. Mouse movements were recorded through general activities of the users in the following papers: \cite{AHMED2007}, \cite{NAKKABI2010}, \cite{AHMED2011}, \cite{FEHER2012} and \cite{SHEN2012}. Mouse activities generated from interaction with applications designed specifically for data collection are used in the following papers: \cite{GAMBOA2004}, \cite{SHEN2013}, \cite{SHEN2014}, \cite{FEHER2012}. Papers \cite{ZHENG2011} and \cite{ZHENG2016} report measurements based on two data sets: one containing the general activities of 30 users using their own computers and one containing web forum usage data of 1000 users.


From the point of view of reproducibility, the first requirement is the availability of the data used for experimentations. If the methods used are accurately described the research results can be reproduced. It is even better if the measurement software is also available. Reproducibility is a desirable property of good research.

The works of Ahmed, Traore, and Nakkabi \cite{AHMED2007}, \cite{NAKKABI2010}, \cite{AHMED2011} are presented in three consecutive papers based on the same data set. Two of these works \cite{AHMED2007}, \cite{AHMED2011} present evaluations on the data set containing data from 22 users, while the third describes the measurement results on an extended data set (48 users). This extended data set is publicly available. The data set does not include raw data, but segmented and processed data. The data are available, the used methods are accurately described, therefore the research is reproducible. We reproduced the results presented in their first paper \cite{AHMED2007}.

The works led by Shen et al. are not reproducible due to the unavailability of their data sets.
We did try to find the data used for their papers and announced as publicly available, but did not find them. They published works based on three data sets collected from: 28 users \cite{SHEN2012}, 37 users \cite{SHEN2013}, and 58 users \cite{SHEN2014}. Works published in \cite{ZHENG2011}, \cite{ZHENG2016}, \cite{GAMBOA2004}, \cite{FEHER2012} are also not reproducible due to the unavailability of the data.


The data segment from which the features are extracted constitutes the modelling unit. The modelling unit may also depend on the type of data collection. 


Gamboa and Fred \cite{GAMBOA2004} used so called strokes for their modelling unit, which are defined as the set of points between two mouse clicks. This unit fits perfectly to the memory game the users had to play with during the data collection. 

Ahmed, Traore and Nakkabi \cite{AHMED2007}, \cite{NAKKABI2010}, \cite{AHMED2011} worked with mouse actions. They used three types of mouse actions: PC - point and click: mouse movement ending in a mouse click; MM - general mouse movement; DD - drag and drop. Histogram-based features were extracted from sequences of mouse actions. The works led by Zheng \cite{ZHENG2011}, \cite{ZHENG2016} used the PC and MM actions defined by Ahmed and Traore \cite{AHMED2007}. Contrary to the feature extraction method presented in \cite{AHMED2007}, the features were extracted from a single action. Feher et al. \cite{FEHER2012} also extracted features from individual mouse actions. However, they used different feature extractors for different action types. 

Firstly, frequent behaviour patterns were identified, then features were extracted in the first work of Shen et al. described in \cite{SHEN2012}. Unfortunately, the feature extraction process is not clearly explained. In a following paper by Shen \cite{SHEN2013}, the features were extracted from a fixed mouse operation task executed 150 times in 15 sessions by each user. Another data collection process was performed by Shen described in paper \cite{SHEN2014}. This time a new task was designed consisting of 8 mouse movements separated by mouse clicks. 13 features were extracted from each mouse movement resulting in a 104-dimensional feature vector.

Chuda and Kratky \cite{CHUDA2014} studied user authentication in web browsing process. They collected mouse usage data from 28 users and reported user authentication accuracy of 87.5\% by using 20 features extracted from mouse data. Unfortunately, the authentication protocol, namely the amount of data used for authentication is not clearly described. Although the data set is not a general purpose one, it is publicly available. In a later paper, Chuda, Kratky and Tvarozek \cite{CHUDA2015} reported user identification in web browsing process using only mouse clicks.  They reported 96\% user identification accuracy on a data set containing data from 20 users by using features extracted from 100 clicks. 

Like Zheng at al. \cite{ZHENG2011}, \cite{ZHENG2016}, Hinbarji et al. \cite{HINBARJI2015}  used mouse movement curves as the basic modelling unit for user authentication. They obtained 9.8\% EER by using features extracted from 100 mouse curves (approximately 5.6 minutes length mouse data). The experiments were conducted on a data set collected from 10 users.

In table \ref{tableRelatedWork} we summarized the important characteristics of the most important existing works on user authentication using mouse dynamics.

\begin{table*}[!htb]
\caption{The most important existing works on mouse dynamics.}
\begin{center}
\begin{tabular}{lllllll}
\toprule
Paper								& \#users	&	Data  &   Data& Features from			    & Main results& Note\\
									& 		 	&         &    availability&			   				&			 &	   \\
\midrule
Gamboa\& Fred, 2004 \cite{GAMBOA2004}	&	50	&	Specific		&	No & stroke		 				& EER: 0.2\% & 200 strokes\\
Ahmed\&Traore, 2007\cite{AHMED2007}		&	22	&	General	& 	Yes& sequence of actions		& EER: 2.46\%&  2000 actions \\
Nakkabi et al., 2010\cite{NAKKABI2010}	&	48	&	General	&	Yes& sequence of actions		& FAR:	0\%, FRR: 0.36\%& 2000 actions \\
Ahmed \& Traore, 2011\cite{AHMED2011} 	&	22	&	General	&	Yes& sequence of actions		& FAR:	0\%, FRR: 0.03\%& 2000 actions \\
Shen et al., 2012\cite{SHEN2012}		&	28	&	General	&	No&	 session& FAR: 0.37\%, FRR: 1.12\%& 3000 mouse \\
                                        &		&			&	  &			&						  & op., 30 min\\

Shen et al., 2013\cite{SHEN2013}		&	37	&	Specific&	No&	 mouse operation &	FAR: 8.74\%, FRR: 7.69\%	&	11.8 sec\\
										&		&			&	  &	 task 			 &								&\\
										
Shen et al., 2014\cite{SHEN2014}		&	58	&	Specific&	No&	 8 consecutive & FAR: 8.81\%, FRR:11.63\%& 6.1 sec\\
										&		&			&	  &	 mouse movements		& 						  & \\

Zheng \& Wang, 2011\cite{ZHENG2011}		&   30 (1000)& General (Specific)&	No&	 action& EER: 1.3\%& 20 clicks  \\
Zheng et al., 2016\cite{ZHENG2016}		&	30 (1000)& General (Specific)&	No&	 action& EER: 1.3\%& 20 clicks  \\
Feher et al., 2012\cite{FEHER2012}						&	25	     & General 			 &	No&	 action& EER: 8.53\%& 30 actions	\\
Hinbarji et al. \cite{HINBARJI2015}		&	10		& General 			& No&	mouse curve& EER: 9.8\%& 100 curves\\
\botrule
\end{tabular}
\label{tableRelatedWork}
\end{center}
\end{table*} 

Mouse dynamics was used not only for authentication, but also for individual performance monitoring and stress detection. Carneiro and Novais \cite{CARNEIRO2012} measured individual performance of the users by their mouse dynamics data.  In another paper, Carneiro et al. studied stress detection during online examinations using mouse movements dynamics \cite{CARNEIRO2015}.

Automatically inferring the gaze position from mouse movements is another research field. Guo and Agichtein \cite{GUO2010} reported 77\% accuracy for predicting gaze position from mouse movements in web searching environment.

\section{The Balabit Mouse Challenge Data Set}

\subsection{Data Set Description}

The Balabit Mouse Challenge data set was published in August 2016. It was used in a Mouse Dynamics data science challenge: "The goal of the challenge was to protect a set of users from the unauthorized usage of their accounts by learning the characteristics of how they use their mouses."

The data set contains two folders, one for the training files and one for the test files. Data were collected from 10 users working over remote desktop clients connected to remote servers. Both training and test data are presented as sessions, however the test sessions are much shorter than the training sessions. The test part of the data set contains both positive (legal) and negative (illegal) sessions.  The number of training and test sessions are shown in Table \ref{table_numsession}. 


\begin{table}
\caption{The number of training and test sessions.}
\begin{center}
{\begin{tabular*}{20pc}{@ {\extracolsep{\fill}} cccc @{} }
\toprule
User	&	training & test    & test \\
        &            & positive& negative\\
\midrule
7		&	7		&	36		&37\\
9		&	7		&	23		&43\\
12		&	7		&	56		&49\\
15		&	6		&	45		&70\\
16		&	6		&	68		&38\\
20		&	7		&	30		&20\\
21		&	7		&	37		&22\\
23		&	6		&	38		&33\\
29		&	7		&	43		&20\\
35		&	5		&	35		&73\\
\midrule
Total	&	65	   &	411	    &405\\
\botrule

\end{tabular*}}{}
\label{table_numsession}
\end{center}
\end{table}

\subsubsection{Raw data}

A session file contains a set of lines and each line represents a recorded mouse event. Each recorded mouse event contains six fields: $(rtime, ctime, button, state, x, y)$. The $rtime$ is the elapsed time in seconds since the start of the session recorded by the network monitoring device. The $ctime$ is also the elapsed time but as recorded by the client computer. The $button$ field represents the current condition of the mouse buttons, and the field $state$ contains additional information about the button. 
The $x$ and $y$ fields are the coordinates of the cursor on the screen (see Fig. \ref{figRawdata}).

Before segmenting the raw data into meaningful units it had to be cleaned. First the duplicated entries were filtered out, then the instances where the mouse was moved outside the remote desktop window (big x, y coordinates) were replaced by previous coordinates.  

The data set also contains recorded mouse scroll events. These events were generated by the scroll button of the mouse. In order to decide whether or not to use these data, we counted the sessions containing this type of events. Since 42\% of the test sessions does not contain scroll events, it was decided not to use this type of raw data.

\subsection{Segmentation}

\begin{figure}[!bt]
\centerline{\includegraphics[width=80mm]{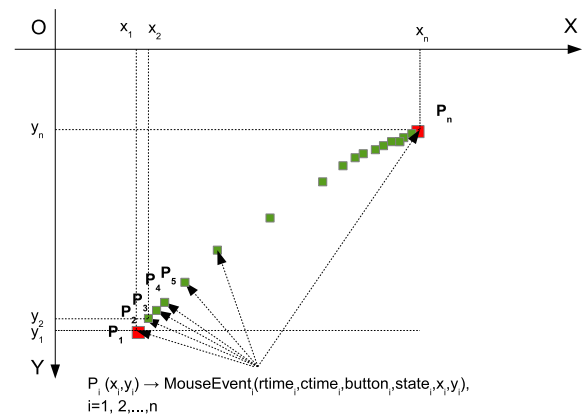}}
\caption{A mouse action: a sequence of consecutive mouse events which describes the movement of the mouse between two screen positions.}
\label{figAction}
\end{figure}

A mouse action is a sequence of consecutive mouse events which describes the movement of the mouse between two screen positions. Since each event contains a screen position, a mouse action which consists of $n$ events is depicted as a sequence of $n$ points: $\{P_1, P_2, \ldots P_n\}$ (see Fig. \ref{figAction}).

We segmented each session's data into three types of mouse actions. These action types were defined by Ahmed and Traore \cite{AHMED2007} as: $\{MM, PC, DD\}$.  $MM$ denotes a general mouse movement action. This type of action simply moves the mouse pointer between two points of the screen. $PC$ denotes a Point Click action. This is similar to the Mouse Movement action, but it ends by a mouse click. $DD$ denotes a Drag and Drop action, which always starts with a left mouse button pressed event, followed by a sequence of mouse drag events, and ends by a left mouse button released event. 

\begin{figure}[!bt]
\centerline{\includegraphics[width=80mm]{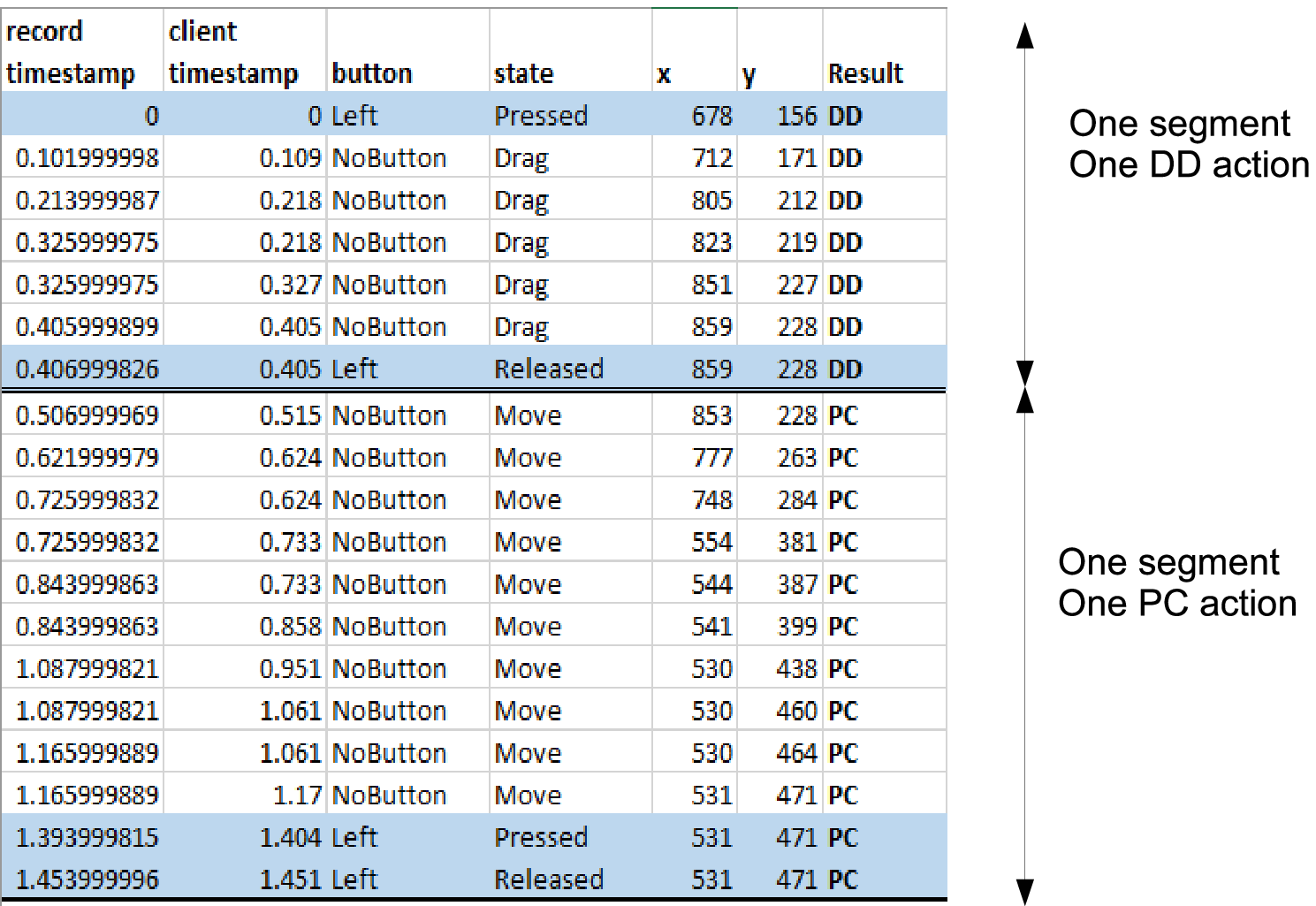}}
\caption{Mouse data segmentation into actions. }
\label{figRawdata}
\end{figure}

Raw data were split into actions in two steps. In the first step we cut the raw data into segments ending in mouse released events. When the previous event of a mouse released event is a mouse pressed event then we have a segment ending in a PC type action, otherwise we have a segment ending in a DD type action (see Fig. \ref{figRawdata}). These segments may contain zero or several MM type actions before the ending in a PC or DD type action. In the second step the MM actions are isolated using a threshold for the time field. We always divide the segment into two actions when the time difference between two consecutive events is higher than a threshold. We used 10 seconds as the time threshold. Finally, short actions having less than 4 events were ignored (spline interpolation requires at least four points).

An example of raw data segmentation into actions can be seen in Fig. \ref{figRawdata}. In this figure two complete segments are shown. Both segments contain only one action. 

\begin{figure*}[!bt]
\centering
\subfloat[training session 5265929106]{\includegraphics[width=.24\linewidth]{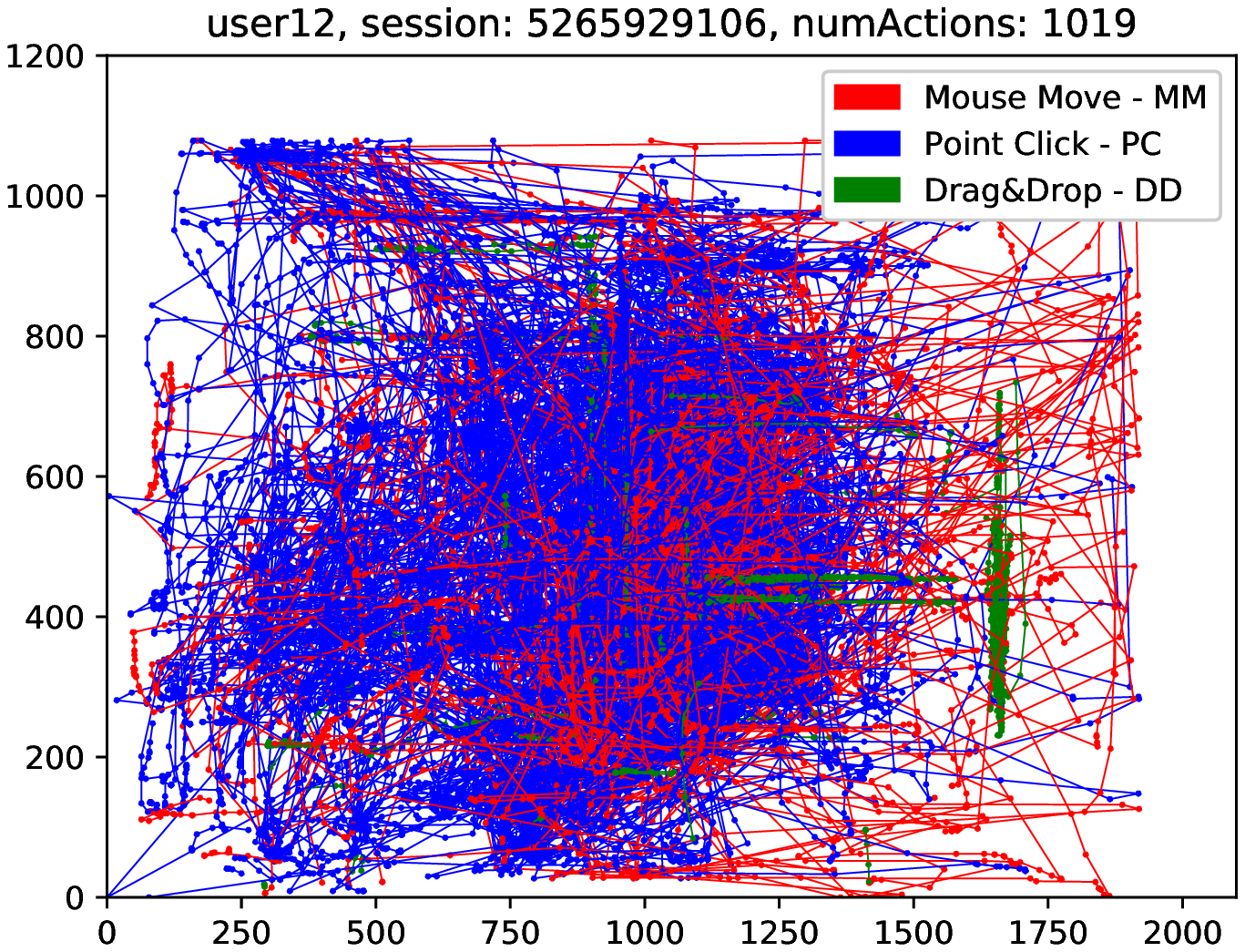}}
\subfloat[test session 0611188910]{\includegraphics[width=.24\linewidth]{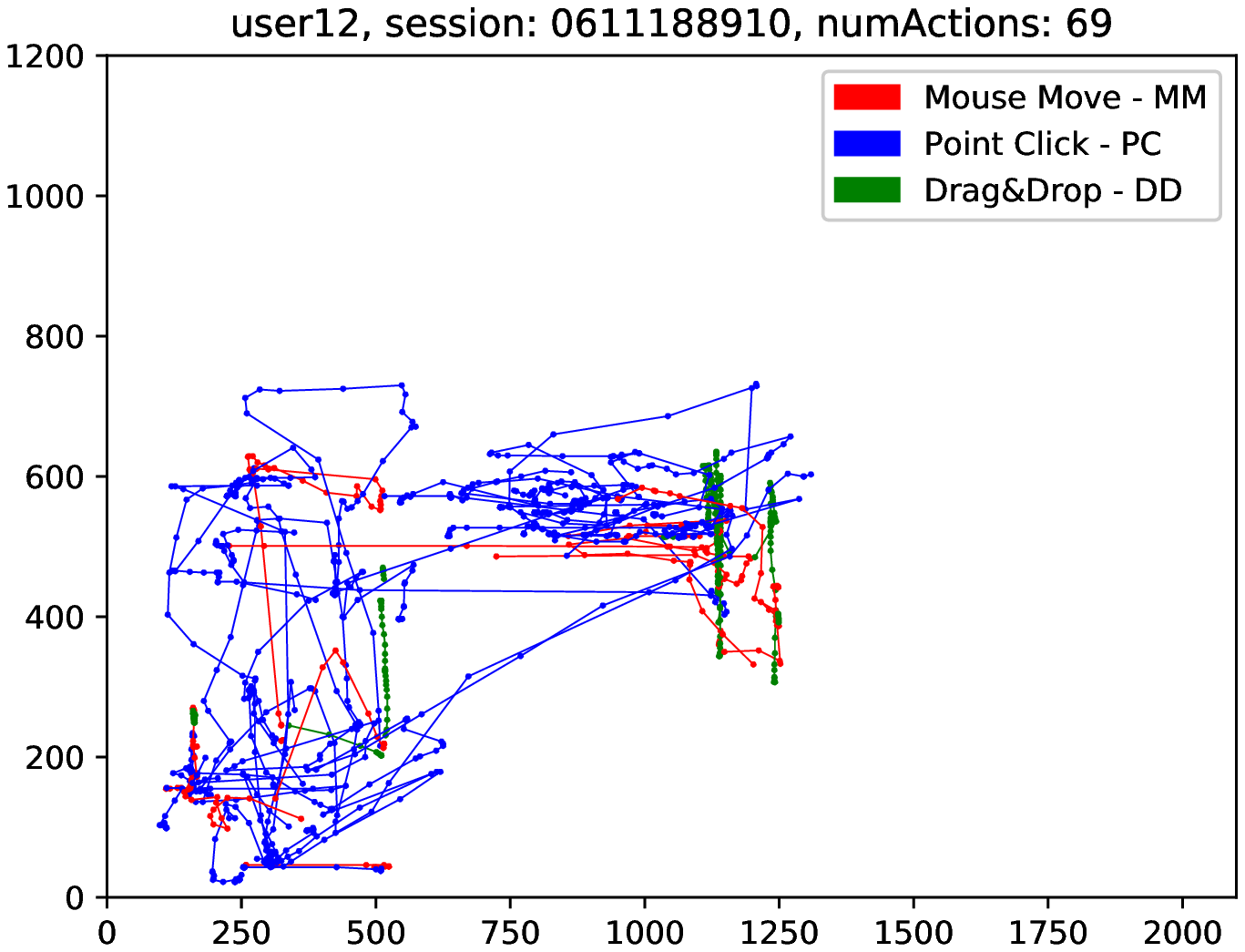}}
\subfloat[test session 1178629549]{\includegraphics[width=.24\linewidth]{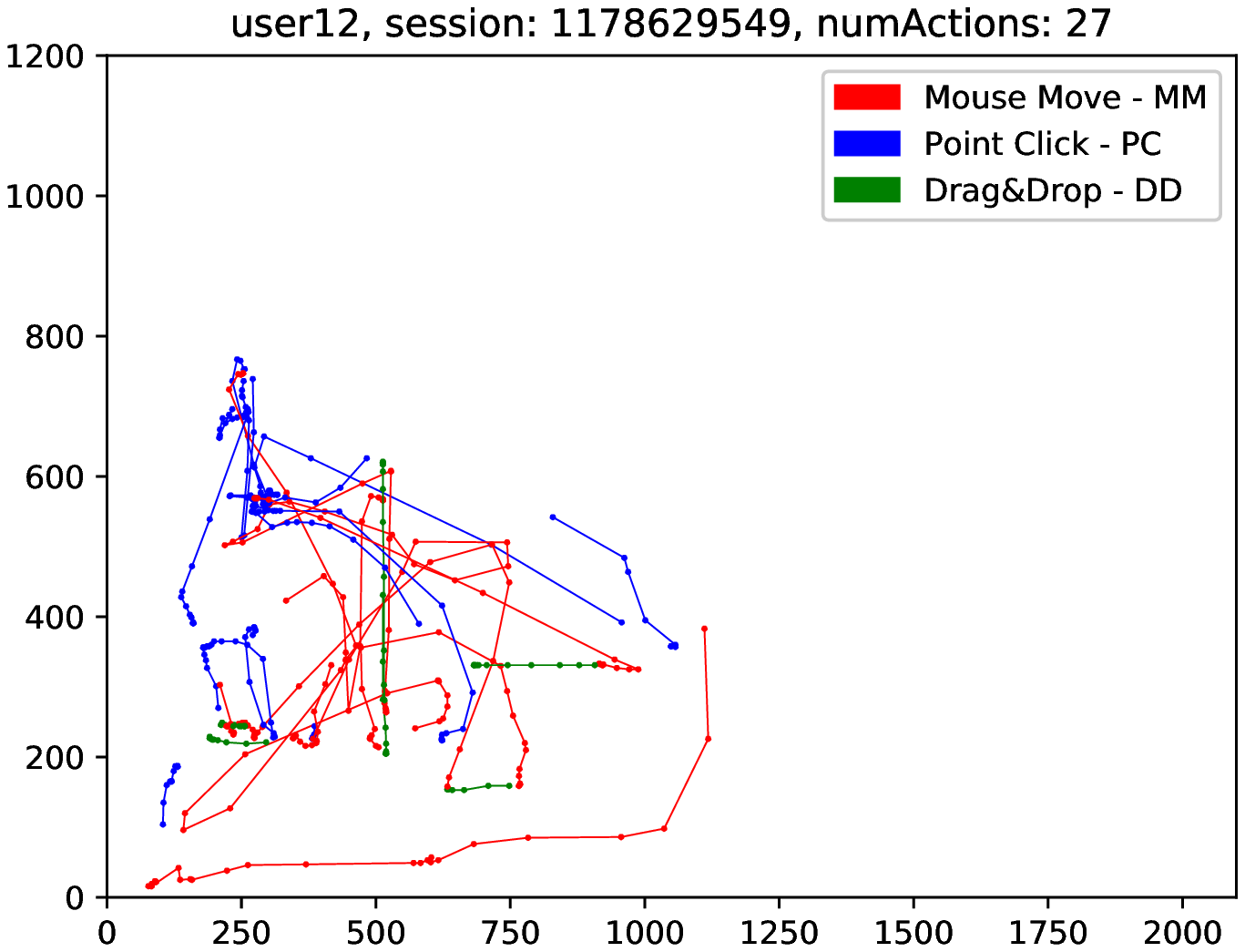}}
\subfloat[test session 9973193301]{\includegraphics[width=.24\linewidth]{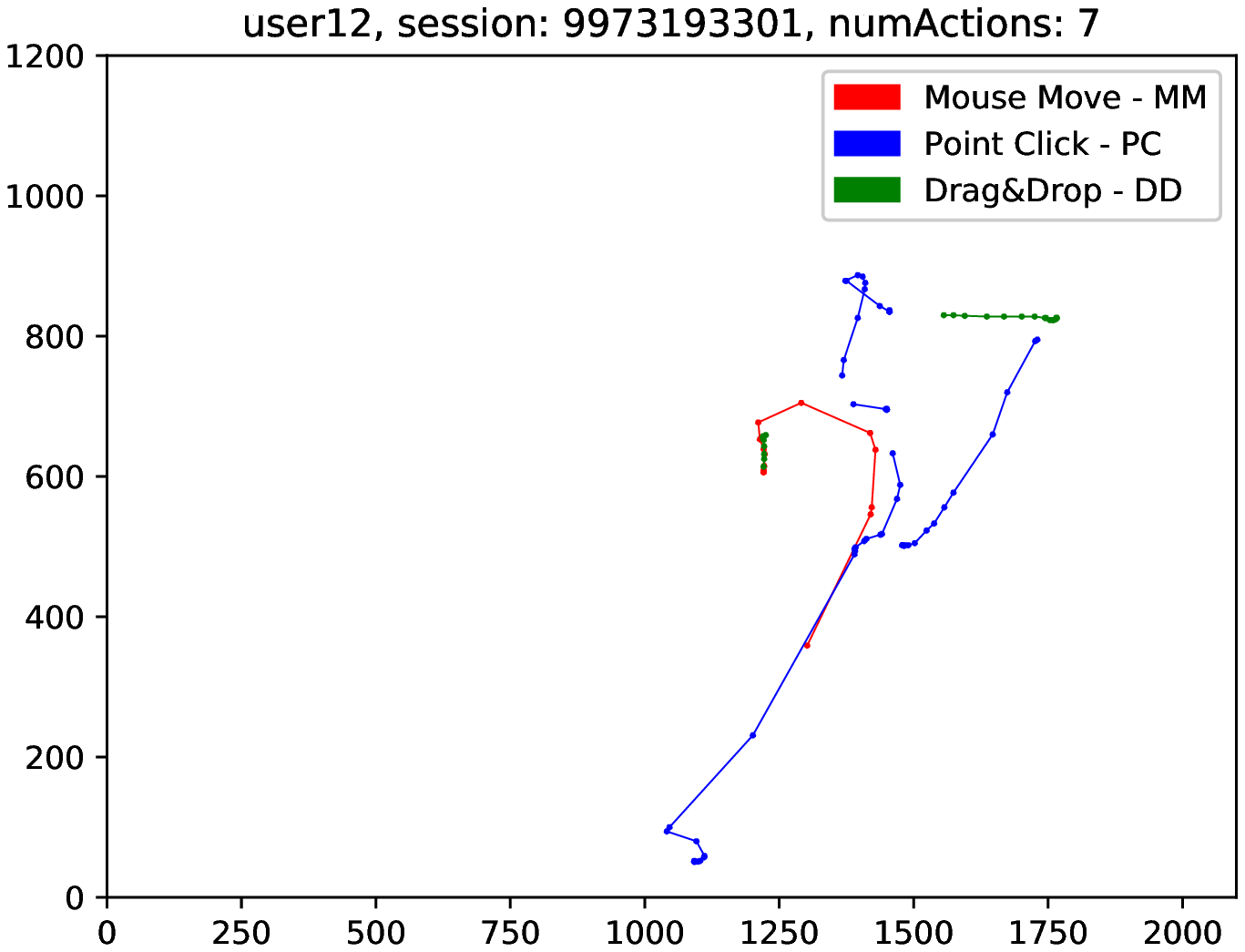}}
\caption{A training and three test sessions of the user 12.}
\label{fig_training_test_rawdata}
\end{figure*}

Fig. \ref{fig_training_test_rawdata} shows the segmented data of four sessions belonging to user12. The first one is a training session and the rest are tests sessions. Note that the training sessions are much longer than the test sessions. 

\subsection{Data Set Statistics}

Table \ref{tableActions} presents the action type statistics for the training and test parts of the data set. It can be seen, that the distribution of the three types of actions is similar in the training and the test parts of the data set. PC actions represents about 70\% of the data, MM actions about 20\% and the remaining 10\% are the DD actions. A training session contains 937 actions on average, at least 237 and at most 2488. Test sessions are much shorter. They contain 50 actions on average, at least 4 and at most 175. This means that even in session-based evaluations for the test part the decision has to be taken based on only 4 actions.

\begin{table}
\caption{Training and test parts statistics} 
\begin{center}

{\begin{tabular*}{20pc}{@ {\extracolsep{\fill}} lccc @{} }
\toprule
			& MM&	PC&	DD\\
\midrule
training&	12828 (21.06\%)&	41610 (68.32\%)&	6467 (10.62\%)\\
test	&	8597 (20.90\%)&		28677 (69.70\%)&	3867 (9.40\%) \\
\botrule
\end{tabular*}}{}
\label{tableActions}
\end{center}
\end{table}

\begin{figure*}[!bt]
\centerline{
\includegraphics[width=.95\linewidth]{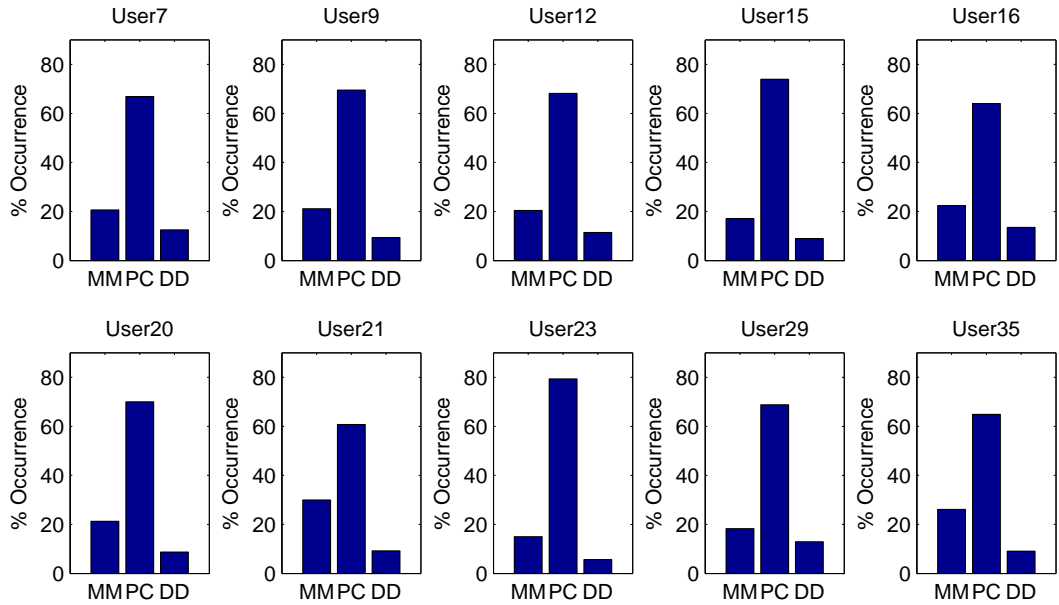}
}
\caption{Histograms of the type of actions for all the 10 users using the training part of the data set. }
\label{figActionType}
\end{figure*}

In order to see the distribution of action types at user level, we computed action type histograms for each user of the data set (see Fig. \ref{figActionType}). For this diagram we used only the training part of the data set. We can see that action type patterns of the users are very similar, despite the fact that during data collection, users performed their individual daily activities.

\section{Proposed User Model}

\begin{figure}[!bt]
\centerline{
\centerline{\includegraphics[width=80mm]{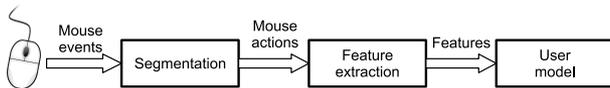}}
}
\caption{User model creation process.}
\label{figArchitecture}
\end{figure}

This section presents the features extracted from mouse actions as well as the machine learning algorithms used for evaluations. Architecture of our user model creation process is shown in Fig. \ref{figArchitecture}.

\subsection{Feature Engineering}
\label{feature_engineering}

We present here the features extracted from individual mouse actions. Similar approaches (using different features) were already studied in the papers \cite{ZHENG2011}, \cite{ZHENG2016}, and \cite{FEHER2012}. Gamboa and Fred \cite{GAMBOA2004} extracted features from action-like units, although they used the term stroke instead of action. 
We have already defined a mouse action as a sequence of consecutive mouse events which describes the movement of the mouse between two screen positions. Even though each mouse event contains six fields (corresponding to a line in the raw data), only three of these are used in the feature extraction: the $x$ and $y$ fields, and the client time stamp $ctime$. The other fields were used in the segmentation step.
Therefore, each action can be represented by a sequence of triplets $(x_i, y_i, t_i), i=1 \ldots n$, where $n$ is the number of mouse events in the action.

Based on the $x$ and $y$ sequences, we computed  $\Theta$, which is the angle of the path tangent with the x-axis calculated using (\ref{eqAngle}):
\begin{equation}
\Theta_i= atan2 \Big{(} \frac {\delta y_i}{\delta x_i} \Big {)}, i=2,\ldots,n; \Theta_1 = 0, 
\label{eqAngle}
\end{equation}

where $\delta y_i= y_i-y_{i-1}$ and  $\delta x_i= x_i-x_{i-1}$. The $atan2$ trigonometric function is a common variation of the standard arctangent function, which produces results in the range $(-\pi,\pi]$.
The following six time series: $v_x$ - horizontal velocity, $v_y$ - vertical velocity, $v$ - velocity, $a$ - acceleration, $j$ - jerk, and  $\omega$ - angular velocity  defined by  (\ref{eqTimeSeries1}) and (\ref{eqTimeSeries2}) were computed based on $x(t)$, $y(t)$, and $\Theta(t)$  series. These six time series were proposed by Gamboa and Fred \cite{GAMBOA2004}:

\begin{equation}
{v_x}_i=\frac {\delta x_i}{\delta t_i},{v_y}_i=\frac {\delta y_i}{\delta t_i},  v_i=\sqrt { {v_x}_i^2+{v_y}_i^2}, i=2,\ldots,n  
\label{eqTimeSeries1}
\end{equation}

where $\delta t_i= t_i - t_{i-1},\quad \delta t_1=0$, ${v_x}_{1}=0, {v_y}_{1}=0, v_{1}=0$ and

\begin{equation}
a_i = \frac {\delta v_i}{\delta t_i}, j_i =\frac{\delta a_i}{\delta t_i}, \omega_i=\frac{\delta \Theta_i}{\delta t_i}, i=2,\ldots,n 
\label{eqTimeSeries2}
\end{equation}

where $a_1=0$, $j_1=0$, and $\omega_1=0$.

The curvature time series $c$, is defined as the  ratio between the angle change and  the travelled distance (\ref{eqTimeSeries3}):

\begin{equation}
c_0=0, \quad c_i = \frac {\delta \Theta_i}{\delta s_i},  i=2,\ldots,n 
\label{eqTimeSeries3}
\end{equation}

where $s_i= \sum_{k=1}^i \sqrt{\delta {x_k} ^2 + \delta {y_k}^2}$ is the length of the trajectory from the starting point of the action to the $i$th point, and $\delta s_i=s_i - s_{i-1}$, $i=2,\ldots, n$ and $s_1=0$.

\begin{figure}[!tb]
\centerline{\includegraphics[width=50mm]{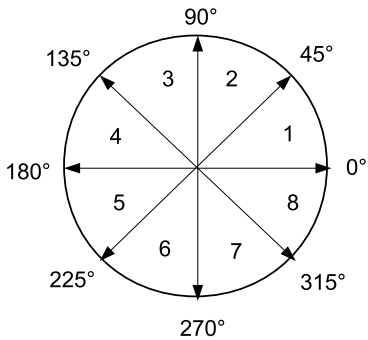}}
\caption{The eight directions. Angles between $0^\circ$ and $45^\circ$ fall into the direction 1.  }
\label{figDirections}
\end{figure}

Mean, standard deviation, and maximal values of the defined six time series were used as features. In addition, we used the following features: the type of the action, the length of the trajectory ($s_n$), the length of the segment between the two endpoints, and the time needed to perform the action. Direction feature is the direction of the end to end line. To reduce the possible direction values, we used the 8 main directions defined by Ahmed and Traore \cite{AHMED2007} (see Fig. \ref{figDirections}). Straightness was computed as the ratio between the segment between the two endpoints of an action and the length of the trajectory. Other used features were: the number of points, which means the number of mouse events contained in an action, the sum of angles, and the largest deviation, which is the largest distance between the points of the trajectory and the segment between the two endpoints. 
Number of sharp angles (less than 0.0005) in the $\Theta$ sequence, and the duration of the first segment of the action  with acceleration $a\_beg\_time$($\sum_{i=2}^{k}{\delta t_i}$, where $k+1$ is the leftmost index where $a_{k+1} \le 0$) were also used as features. Table \ref{tableActionFeatures} shows all the features extracted from a mouse action.

\begin{table}
\caption{Features extracted from a single mouse action}
\begin{center}
{\begin{tabular*}{20pc}{@ {\extracolsep{\fill}} lll @{} }
\toprule
Name&	Description& \#features\\
\midrule
$v_x$		& mean, std, min, max&	4\\
$v_y$		& mean, std, min, max&	4\\
$v$			& mean, std, min, max&	4\\
$a$			& mean, std, min, max&	4\\
$j$			& mean, std, min, max&	4\\
$\omega$	& mean, std, min, max&	4\\
$c$			& mean, std, min, max&	4\\
type				& {MM, PC, DD} 			  		&	1\\
elapsed time		& $t_n - t_1$					&	1\\
trajectory length  	& $s_n$ 						&	1\\
dist\_end\_to\_end	& $|P_1P_n|$					&	1\\
direction			& see Fig.\ref{figDirections}	& 	1\\
straightness		& $ \frac{|P_1P_n|}{s_n}$			&	1\\
num\_points			& $n$							&	1\\
sum\_of\_angles		& $\sum_{i=1}^{n} \Theta_i$		&	1\\
largest\_deviation	& $max_i \{d(P_i,|P_1P_n|)\}$		&	1\\

sharp angles		& \# $\{\Theta_i|\Theta_i<TH\}$		&	1\\
a\_beg\_time		& accel. time at the beginning	& 	1\\
\midrule
Total				&								& 	39\\
\botrule
\end{tabular*}}
\label{tableActionFeatures}
\end{center}
\end{table}

\begin{figure*}[!bt]
\centering
\includegraphics[width=.95\linewidth]{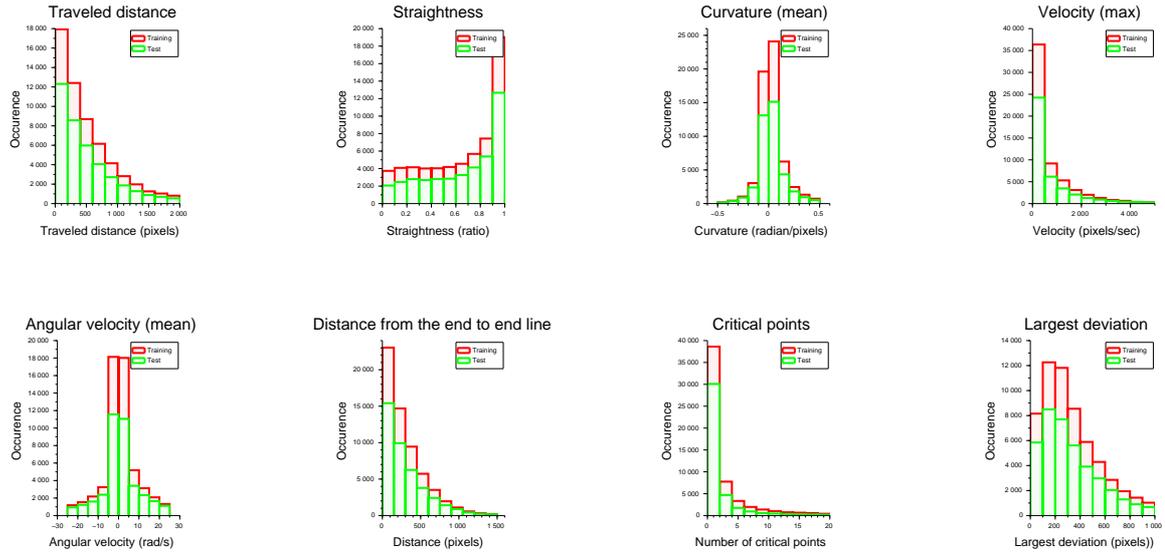}
\caption{The distribution of different features in the training and the test parts of the data set.}
\label{fig_histo_feat1}
\end{figure*}

The distribution of different features in the training and the test parts of the data set are shown in Fig. \ref{fig_histo_feat1}. It can be seen, that the distribution of features in the training and the test parts are similar. 
The travelled distance histogram shows the frequency of actions within different distance ranges. Usually, users performed actions with short travelled distances more frequently than with long ones. In the straightness histogram we can see that there are many relatively straight actions in both the test and the training parts of the data set. This is also confirmed by the histogram of the curvature of the actions. The velocity histogram shows that users prefer low speed movements. Angular velocity histograms reveal the tendency of the users making mostly straight mouse movements. The histogram of the distance between the two endpoints is similar to that of the travelled distance. Users make mouse movements having a low number of critical points. This is consistent with straightness, curvature and largest deviation histograms.

\subsection{User Profiles}

It is known that in some types of behavioural biometrics binary or two-class classifiers perform better than one-class classifiers \cite{ANTAL2015}. Using information about other users' behaviour results in better user profiles. However, it is not always possible to collect negative data (e.g. keystroke dynamics,  password typing rhythm). Fortunately, in mouse dynamics this is possible, therefore we used binary classifiers. 

For each user we created a binary classifier which is responsible for deciding whether some amount of mouse data belong or not to a given user. Several classifiers integrated into the Weka Machine learning toolkit \cite{WEKA} were tried and Random forest  algorithm \cite{RANDOMFOREST2001} proved to be the best. This algorithm performs well even in the case of correlated features. Therefore, we trained a separate Random forest  classifier for each user. These classifiers were trained by using balanced training sets. 

Training sets contain two classes: the genuine (positive) and the impostor (negative). For each user we selected a user-specific data set having an equal amount of genuine and impostor mouse actions in order to avoid bias. A user having $N_i$ genuine mouse actions would have in its user-specific data set these $N_i$ genuine actions and $N_i$ impostor actions. Impostor actions were selected from the other users. Having a total number of $n$ users, for each user-specific data set we have one genuine user (positive data samples) and $n-1$ impostor users (negative data samples). Consequently, $N_i/(n-1)$ impostor actions were selected from each impostor user.

After the training stage, the created classifiers were evaluated. Two types of evaluations were used: action-based and set of actions-based. In the action-based evaluation the classifier output is a score which represents the probability of the action belonging to the positive class. However, in the case of set of actions based evaluation, we need the decision to be taken based on all the actions in the set. In the following we explain the score computation for this case. Let $A=\{a_1,a_2,\ldots,a_k\}$ be a set of actions. For each action we computed the probability of belonging to the positive class. This resulted in a set of $k$ probabilities: $\{p_1,p_2,\ldots,p_k\}$, then we computed a single score for the set of actions as the average of the actions probabilities (eq. \ref{eq:probability}).
\begin{equation} 
P(A) = \frac {1}{k} \sum _{i=1}^{k}{p_i}
\label{eq:probability}
\end{equation}

\section{Experimental Results}
\label{Experimental Results}

\subsection{Evaluation Metrics}
\label{evaluation_measures}

We report evaluation results in terms of classification accuracy (ACC), area under curve (AUC), false negative rates (FNR), false positive rates (FPR), and equal error rates (EER). Besides these values, we report the performance of our detection system by ROC curves \cite{FAWCETT2006}. ROC curves are especially meaningful for biometrics because they show the performance of the biometric system on several operating points corresponding to different decisional thresholds.

Accuracy is defined as the percentage of correctly classified instances. In the case of ROC curves, first, we computed a positive score list using the positive test data and a negative score list using the negative test data. Then, based on these two lists the false positive rates (FPR) and false negative rates (FNR) were computed for different threshold values ($threshold \in [0,1]$). FPR  is the ratio of the number of negative samples accepted as positive divided by the total number of negative samples (false positive rate). FNR is the ratio of the number of positive samples rejected divided by the total number of positive samples (false negative rate).  EER values express the error rate for a binary classifier detection threshold for which the FPR  and the FNR are equal. The AUC value is computed using the ROC curve which plots false positive rates against true positive rates: it is the area under the ROC curve. "The AUC of a classifier is equivalent to the probability that the classifier will rank a randomly chosen positive instance higher than a randomly chosen negative instance." \cite{FAWCETT2006} AUC is often used to compare classifiers.

All the evaluations were performed with a Java application using Weka Machine Learning toolkit (version 3.8). 

\subsection{Evaluation scenarios}

The Balabit data set has two parts: a training and a test part. We are going to present two evaluation scenarios: (A) both the training and the testing were performed using exclusively the training part of the data, (B) the training had been performed on the training part of the data set and the testing on the test part of the data set. 

In both scenarios we evaluated the binary classifiers using decisions: (i) based on single action, (ii) based on a set of actions (or sequence of actions). Evaluation based on a set of actions are also referred as periodic authentication \cite{MONDAL2015} because the users can always perform a fixed number of actions before the system checks their identity. 

\subsection{Scenario A: Balabit data set -- using only the training part}

In this subsection we describe the details of experiments based on the training part of the Balabit data set. This part of the data set contains 60905 actions from 65 sessions. 

\subsubsection{Action-based evaluation}

A binary classifier was built from the user-specific data set selected as explained earlier. These data sets were evaluated using Random forest classifier and 10-fold cross-validation. Table \ref{tableUserAccuracyTraining} reports detailed results for each of the 10 users of the data set. Besides the user-specific evaluation metrics, the mean and the standard deviation are also reported. Using 10-fold cross-validation 80,17\% average accuracy and 0,87 average AUC was obtained. We should note that each classifier was evaluated separately. This means that user specific metrics were computed based on user-specific decision thresholds. We report the average and the standard deviation of these values. It is important to note that high accuracies (over 93\%) and also high AUC values (over 0.97) were obtained for two users (7 and 9), and lower accuracies (between 72\% and 80\%) and also lower AUC values (between 0.80 and 0.89) for the other 8 users. Except for a few users (7, 9 and 29), the false positive rates are much higher than the false negative rates. This means that using only one mouse action a lot of impostors are accepted as they were genuine users.

\begin{table}[!b]
\caption{Scenario A: Action-based evaluation}

\begin{center}
{\begin{tabular*}{20pc}{@ {\extracolsep{\fill}} lrrrrr @{} }
\toprule
User&	ACC& AUC&			FNR&FPR\\
\midrule
7	&93.27&0.9725&		0.0138&	0.1208\\
9	&97.41&0.9975&		0.0132&	0.0386\\
12	&77.17&0.8560&		0.1835&	0.2736\\
15	&77.43&0.8621&		0.1642&	0.2873\\
16	&72.69&0.8132&		0.1648& 0.3814\\
20	&78.54&0.8731&		0.1759&	0.2532\\
21	&76.65&0.8462&		0.1739&	0.2931\\
23	&75.74&0.8414&		0.1960&	0.2892\\
29	&80.50&0.8954&		0.2017&	0.1881\\
35	&72.01&0.8010&		0.1967&	0.3632\\
\midrule 
Avg& 80.17&0.8761&		0.1484&	0.2488\\
Std.& 8.45&0.0637&		0.0723&	0.1057\\
\botrule
\end{tabular*}}
\end{center}
\label{tableUserAccuracyTraining}.
\end{table}

\subsubsection{Which type of action is the most user specific?}

The training data were separated into three sets, each set containing only MM, PC, or DD actions respectively. In order to see which of the three types of actions is the most user specific we repeated the action-based evaluation three times using the corresponding training set.  Detailed results are shown in Table \ref{Table_MM_PC_DD}. It can be seen that the results are very similar for MM and PC actions. This is not a surprising result because the only difference between the two type of actions is that PC actions contains a mouse click event also. Even though DD type actions are 10\% of the data set, if only this type of action is used for both model building and testing, the results are 3\% better than for using only PC or only MM type actions.

\begin{table}[!b]
\caption{Scenario A: Using only one type of action.}

\begin{center}
{\begin{tabular*}{20pc}{@ {\extracolsep{\fill}} lrrrrrr @{} }
\toprule
		&\multicolumn{2}{c} {MM} & \multicolumn{2}{c}{PC}& \multicolumn{2}{c}{DD}\\
\midrule		
User&	ACC& AUC& ACC& AUC& ACC& AUC\\
\midrule
7&	91.50&	0.9576&	93.69&	0.9764&	89.86&	0.9569\\
9&	95.62&	0.9953&	97.65&	0.9984&	94.84&	0.9902\\
12&	75.00&	0.8465&	76.85&	0.8504&	84.28&	0.9240\\
15&	78.58&	0.8599&	76.29&	0.8513&	83.77&	0.9230\\
16&	72.20&	0.8041&	72.26&	0.8001&	78.00&	0.8644\\
20&	75.71&	0.8334&	78.09&	0.8638&	81.89&	0.9071\\
21&	75.53&	0.8328&	75.82&	0.8413&	75.94&	0.8386\\
23&	73.97&	0.8152&	74.54&	0.8302&	79.32&	0.8725\\
29&	78.05&	0.8679&	79.61&	0.8889&	85.92&	0.9428\\
35&	72.23&	0.7928&	71.71&	0.7970&	73.59&	0.8290\\
\midrule
Avg.&	78.84&	0.8605&	79.65&	0.8698&	\textbf{82.74}&	\textbf{0.9048}\\
Std.&	8.09&	0.0660&	8.83&	0.0679&	\textbf{6.46} &	\textbf{0.0527}\\
\botrule
\end{tabular*}}
\end{center}
\label{Table_MM_PC_DD}.
\end{table}

\subsubsection{Set of actions-based  evaluation}

In this case scores were computed for sets of actions varying the number of actions in a set. In this case, we report the performance for the whole system. This means that both AUC and EER were computed based on a global threshold, which was computed using a common positive and negative score lists for all users of the data set. Table \ref{tableTrainingAUCEER} shows the AUC and EER values as a function of number of actions. It can be seen that after 13 actions the AUC became 1, however for an EER of 0.0004 (0.04\%) 20 actions were necessary.

\begin{table}[!b]
\caption{Scenario A: Set of actions-based evaluations.}

\begin{center}
{\begin{tabular*}{20pc}{@ {\extracolsep{\fill}} lrr@{} }

\toprule	
\#actions&	AUC&	EER\\
\midrule
1&	0.8806&	0.2128\\
2&	0.9490&	0.1274\\
3&	0.9763&	0.0808\\
4&	0.9883&	0.0545\\
5&	0.9940&	0.0375\\
6&	0.9968&	0.0269\\
7&	0.9982&	0.0186\\
8&	0.9991&	0.0136\\
9&	0.9995&	0.0096\\
10&	0.9997&	0.0069\\
11&	0.9998&	0.0051\\
12&	0.9999&	0.0039\\
13&	1.0000&	0.0026\\
14&	1.0000&	0.0018\\
15&	1.0000&	0.0017\\
16&	1.0000&	0.0016\\
17&	1.0000&	0.0012\\
18&	1.0000&	0.0014\\
19&	1.0000&	0.0007\\
20&	1.0000&	0.0004\\
\botrule
\end{tabular*}}
\end{center}
\label{tableTrainingAUCEER}.
\end{table}

\subsubsection{Informativeness of the features}

Informativeness of the features has also been evaluated using the GainRatioAttributeEval evaluator from the Weka Machine Learning toolkit (version 3.8). This evaluator computes the gain ratio with respect to the class for each feature. Larges gain ratios indicate better features. Features ranking is shown in Table \ref{tableFeatureInformativenessSingleAction}. It can be seen that the most informative feature was the acceleration at the beginning of the mouse action followed by the number of critical points. Jerk related features were also found to be more informative than velocity related features. The least informative  feature was the direction of the action.

\begin{table*}[!htb]
\caption{Informativeness of the features. }
\begin{center}
{\begin{tabular*}{20pc}{@ {\extracolsep{\fill}} llrllrllr @{} }
\toprule
Ranking& Feature&	Gain ratio&	Ranking& Feature&Gain ratio& Ranking& Feature&	Gain ratio\\
\midrule
  1&	a\_beg\_time& 		0.13422&			14& max\_curv&			0.06654&	27&	min\_v&					0.02255\\									
  2&	num\_critical\_points& 0.12266& 		15&	min\_vy&			0.06428&	28&	sd\_vy&					0.02223\\								  	
  3&	min\_jerk&			0.09648& 			16&	min\_a&				0.06339& 	29&	sd\_vx&					0.02141\\								 	
  4&	num\_points&		0.09458&  			17&	max\_vx&			0.06182&	30&	mean\_omega&			0.02093\\ 		  
  5&	min\_omega&			0.09401&  			18&	min\_vx&			0.06071&  	31&	straightness&			0.02002\\		
  6&	max\_omega&			0.09292&   			19&	max\_a&				0.0522& 	32&	travelled\_distance\_pixel&	0.01982\\  	
  7&	mean\_jerk&			0.08626&	 		20&	sd\_a&				0.05038&  	33&	dist\_end\_to\_end\_line&	0.01751\\ 	
  8&	sd\_jerk&			0.08089& 	 		21&	sd\_curv&			0.03538& 	34&	mean\_v&	0.01732 \\   	
  9&	min\_curv&			0.07969&  			22&	elapsed\_time&		0.03416&	35&	mean\_curv&	0.01396 \\		
 10&	mean\_a&			0.0774&    			23&	max\_v&				0.03385&    36&	mean\_vx&	0.01154 \\
 11&	max\_jerk&			0.0754&				24&	sd\_omega&			0.03367&	37&	mean\_vy&	0.01121 \\
 12&	sum\_of\_angles&	0.0743&				25&	largest\_deviation&	0.0295&		38&	type\_of\_action&	0.00996\\
 13& 	max\_vy&			0.06765& 			26&	sd\_v&				0.02501&	39&	direction\_of\_movement&	0.00291\\
 
\botrule
\end{tabular*}}
\end{center}
\label{tableFeatureInformativenessSingleAction}.
\end{table*}

\subsection{Scenario B: Balabit data set -- using both training and test parts}

In this subsection we describe the details of the evaluations based on the whole Balabit data set. The main difference compared to scenario A is that, in this section the training part of the data set was used only for building the classifiers, and all the testings were performed exclusively on the test part of the data set.

\subsubsection{Action-based evaluation}

The binary classifiers were built similarly to the scenario A, except that we used all the data in the training part for training the classifiers. Having the trained classifiers,  we used the test data (both positive and negative sessions) and evaluated the classifiers. The detailed results are reported in Table \ref{tableUserAccuracyTest}. Compared to the same type of evaluation used in scenario A (see Table \ref{tableUserAccuracyTraining}), we observed a decrease of 10\% in terms of AUC and 8\% in terms of accuracy. In terms of errors,  in the case of FNR we observed a small increase, but the  FPR almost doubled. 

\begin{table}[!b]
\caption{Scenario B: Action-based evaluation}

\begin{center}
{\begin{tabular*}{20pc}{@ {\extracolsep{\fill}} lrrrrrr @{} }
\toprule
User&	ACC& AUC& FNR& FPR\\
\midrule
7	&98.63&0.9987&	0.0128&	0.0146\\	
9	&99.16&0.9997&	0.0170& 0.0038\\	
12	&68.61&0.7666&	0.2137&	0.4167\\	
15	&53.20&0.6293&	0.2729&	0.5777\\	
16	&63.54&0.6226&	0.1854&	0.6917\\	
20	&82.68&0.9192&	0.1522&	0.2053\\
21	&66.54&0.6676&	0.1963&	0.5906\\
23	&65.64&0.7270&	0.2298&	0.4434\\
29	&70.96&0.7389&	0.2260&	0.4511\\
35	&53.94&0.6761&	0.1708&	0.5976\\
\midrule
Avg.&72.29&0.7746&	0.1677&	0.3992\\
Std.&16.30&0.1456&	0.0873&	0.2450\\
\botrule
\end{tabular*}}
\end{center}
\label{tableUserAccuracyTest}.
\end{table}

\subsubsection{Set of actions-based evaluation}

In this case, scores were computed for sets of actions varying the number of actions in a set. We report the AUC and EER values as a function of the number of actions. We note that test sessions contain a varying number of actions (between 4 and 175, on average 50). Let us denote $m$ the number of actions contained in a test session, and $n$ the number of actions contained in a set of actions. Then, for this session we computed $m/n$ scores, one for each n-length set of action. Sessions containing less than $n$ actions were ignored. Table \ref{tableTestAUCEER} shows the AUC and EER values as a function of number of actions. It is worth comparing table \ref{tableTestAUCEER}  to table \ref{tableTrainingAUCEER}. We can observe the same tendency as for action-based evaluation. AUC values are lower with 10\% compared to the same type of evaluation in scenario A. The best AUC is 0.89 and the corresponding EER is 0.18.

\begin{table}[!b]
\caption{Scenario B: Set of actions-based evaluations.}
\begin{center}
{\begin{tabular*}{20pc}{@ {\extracolsep{\fill}} lrr@{} }
\toprule	
\#actions&	AUC&	EER\\
\midrule
1&	0.7864&	0.3024\\
2&	0.8214&	0.2684\\
3&	0.8392&	0.2526\\
4&	0.8512&	0.2415\\
5&	0.8600&	0.2342\\
6&	0.8667&	0.2277\\
7&	0.8727&	0.2271\\
8&	0.8759&	0.2168\\
9&	0.8792&	0.2114\\
10&	0.8838&	0.2086\\
11&	0.8857&	0.2051\\
12&	0.8863&	0.2061\\
13&	0.8914&	0.1953\\
14&	0.8908&	0.1951\\
15&	0.8932&	0.1924\\
16&	0.8943&	0.1944\\
17&	0.8963&	0.1914\\
18&	0.8964&	0.1952\\
19&	0.8983&	0.1897\\
20&	0.8994&	0.1880\\
\botrule
\end{tabular*}}
\end{center}
\label{tableTestAUCEER}.
\end{table}

\subsubsection{Session-based evaluation}

This evaluation is similar to the set of actions-based one. In this case we used also a set of actions for decision: all the actions belonging to a test session. For each session we computed exactly one score. The number of scores obtained is equal to the number of test sessions (816).

\begin{figure*}[!bt]
\centering
\subfloat[ROC curves]{\includegraphics[width=.33\linewidth]{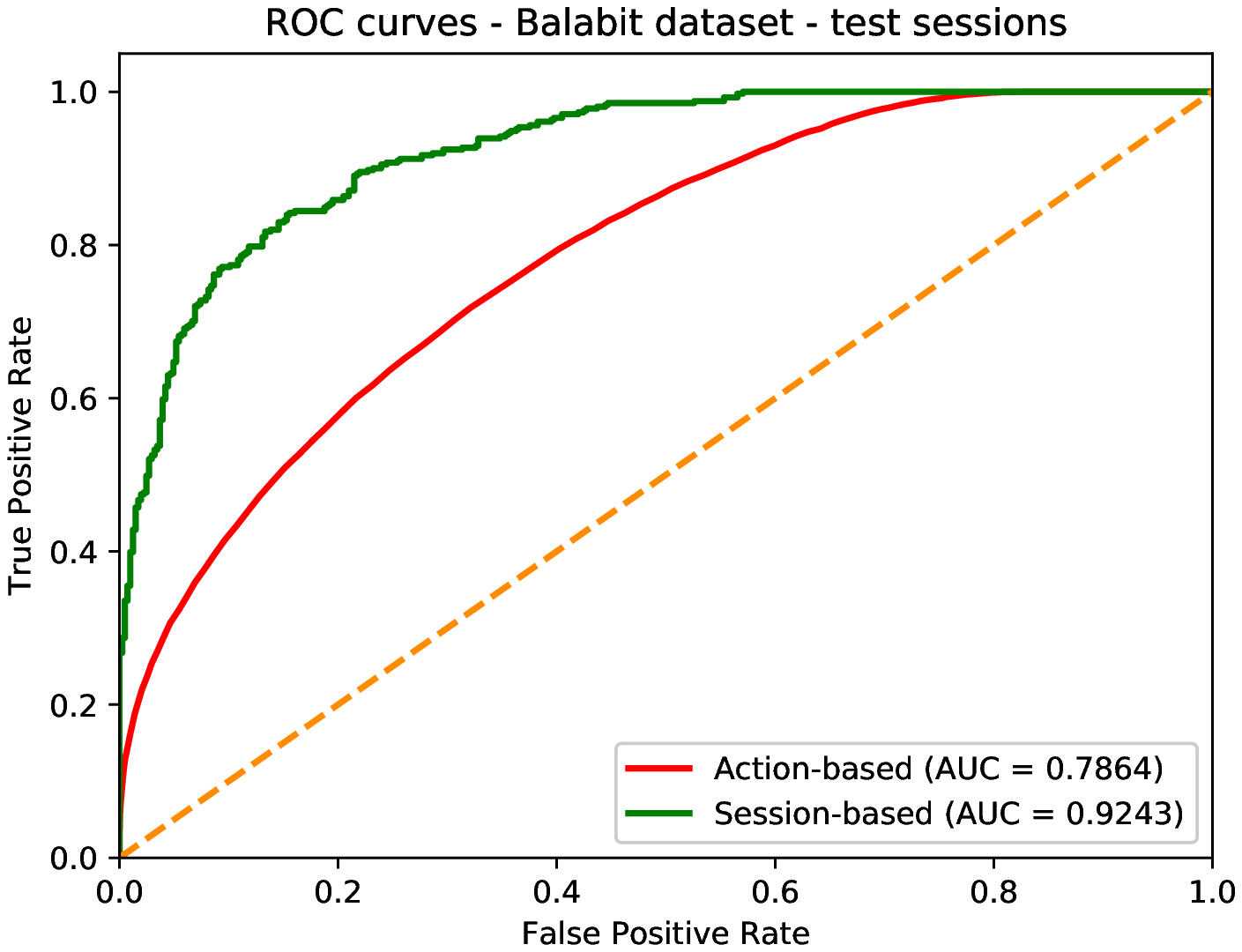}	}
\subfloat[Score distribution -- Action-based evaluation]{\includegraphics[width=.33\linewidth]{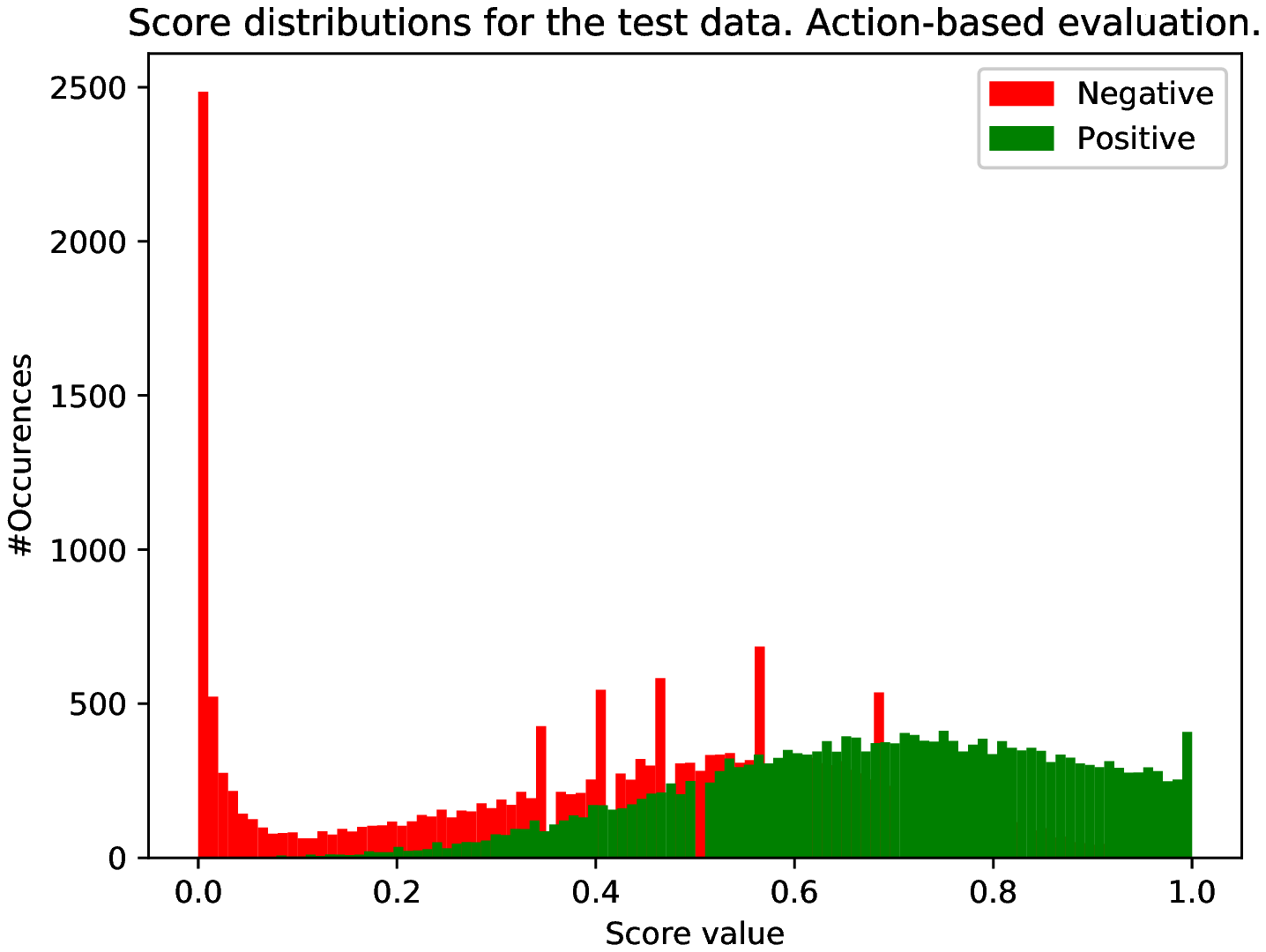}}
\subfloat[Score distribution -- Session-based evaluation]{\includegraphics[width=.33\linewidth]{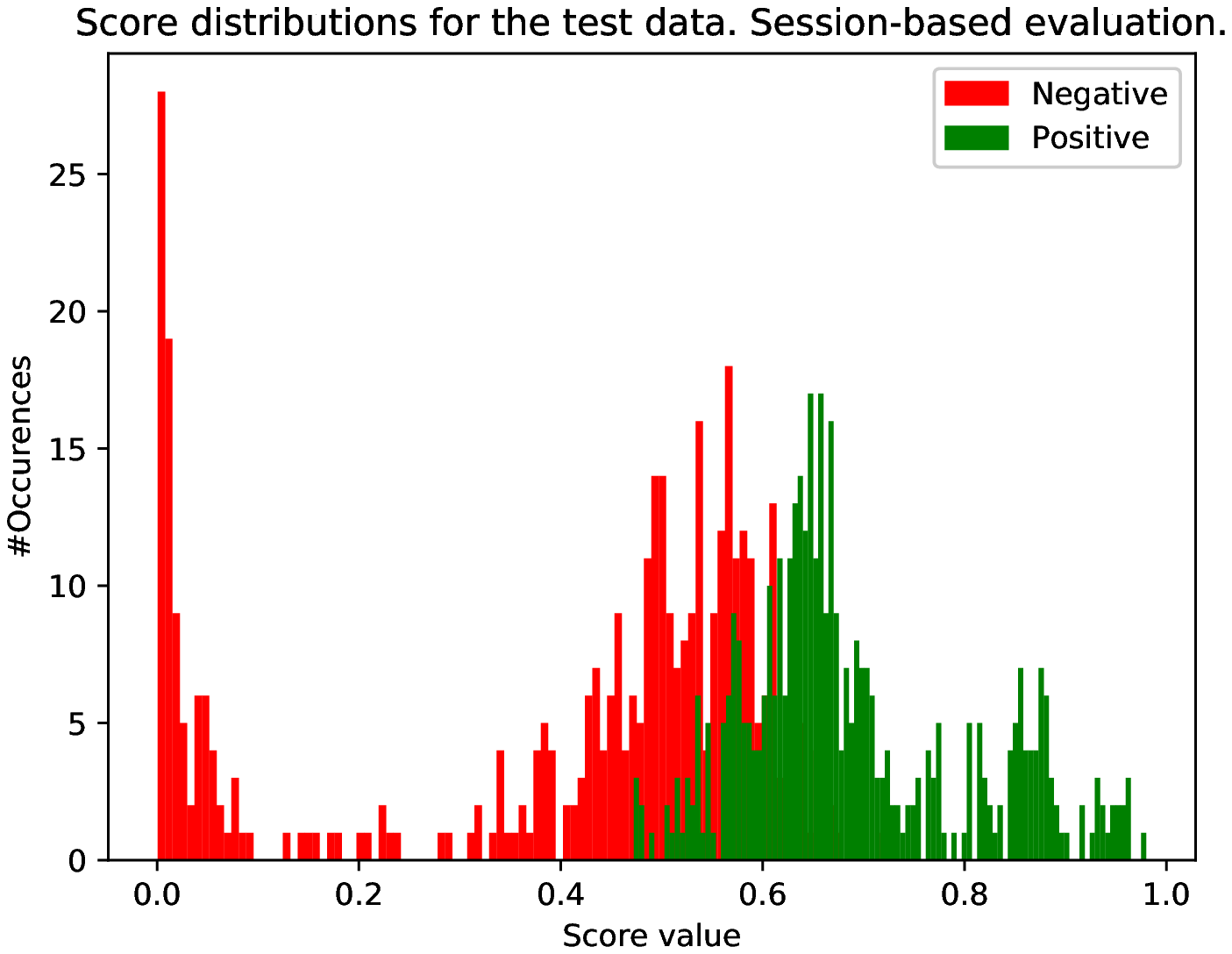}}
\caption{ROC curves. Action-based vs. session-based evaluations.}
\label{figROC_case2}
\end{figure*}

Scores were computed for all positive and negative test sessions, which resulted in a positive and negative score lists. The distribution of these scores and its corresponding ROC curve is shown in Fig. \ref{figROC_case2}. For comparison we included the ROC curve obtained for action-based evaluation and the corresponding scores' distribution too. We observe an increase of 14\% in terms of AUC when using session-based evaluation.

\subsubsection{Effect of Signal Smoothing}

\begin{figure}[!tb]
\centerline{\includegraphics[width=1\linewidth]{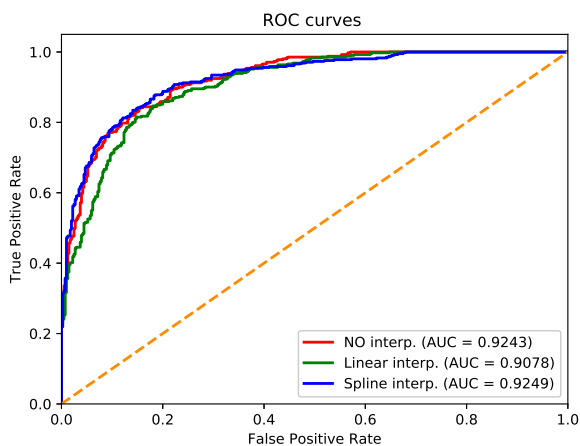}}
\caption{Effect of signal smoothing. Evaluation: session-based.}
\label{figROCSmooth}
\end{figure}

In order to smooth the unevenly (event-based) sampled signal, we interpolated the mouse movements to temporally homogeneous samples (the smoothed curve was re-sampled with a frequency of 20 Hz). Both linear and spline interpolation were applied to the signal. We report results for scenario B, session based evaluation. The results are depicted on Fig. \ref{figROCSmooth}. The spline interpolation did not change the result significantly, but the linear interpolation resulted in a 2\% decrease of the AUC value.

\section{Discussion}

Gamboa and Fred \cite{GAMBOA2004} demonstrated experimentally that mouse data collected through a memory game (all the users working with the same application) can be used for user authentication by achieving 0.2\% EER for 200 mouse strokes (a kind of mouse action). Ahmed and Traore \cite{AHMED2007} extended Gamboa and Fred's work by collecting mouse data resulting from a general computer usage scenario (all the users performing their daily work). They obtained 2.46\% EER by using 2000 actions for authentication. Despite the good results, the high number of actions required to verify a user's identity makes their method impractical for access control systems.

Compared to these two works our result of 0.04\% EER obtained for 20 actions is very good. We should note that this very good result was obtained only for the scenario A (evaluation based on the training part of the data set). More modest result of 18.80\% EER (20 actions) was obtained for the test part of the data set.
 
Like Feher et al. \cite{FEHER2012}, we evaluated methods based on both single actions and aggregation of actions (set of actions). They reported the best EER for using 30 actions for authentication. Our findings are similar for the Balabit data set:  using more than 15 actions, the EER and the AUC values did not improve significantly.


We reported our findings related to the Balabit data set which is quite a challenging one due to the short test sessions. Besides the good detection results we were particularly attentive to the reproducibility of our measurements. The results may not be scalable to a larger system, because it contains data only from 10 users. There is still need for a larger mouse dynamics data set.

\section{Conclusions}

In this paper we have presented an intruder detection study using the Balabit mouse dynamics data set. This data set was specially developed for intruder detection. We especially took care of the reproducibility of our work. Both the data set and the software are publicly available.

The data set was evaluated using two scenarios: (A) using only the training part of the data set, and (B) using the whole data set. Scenario B proved to be quite challenging due to the short test sessions. In both scenarios we used binary classifiers. These binary classifiers were evaluated based on single action and set of actions. 

We have seen that one action does not contain enough information about the user mouse usage behaviour, therefore more actions are required for an accurate decision. Scenario A resulted in almost perfect classification using 20 actions for detection (AUC: 1, EER: 0.04\%), however in scenario B more modest results were obtained for the same number of actions (AUC: 0.89, EER: 18.80\%). A slight improvement was obtained by session based evaluation (0.92 AUC). In this case the sessions contained 50 actions on average, at least 4 and at most 175.

Action specific classifiers were trained using only one type of mouse actions. Despite the fact that drag and drop actions constitute only 10\% of the data set, these proved to be the most user specific action. 
  
Both linear and spline interpolation were applied to the unevenly sampled mouse data. The interpolation did not improve the intruder detection system accuracy.

Our findings suggest that mouse dynamics can be considered in an intruder detection system along with other behavioural biometrics. Nevertheless, one should have enough test data in order to perform the detection with a good accuracy. We chose the Balabit data set because to the best of our knowledge it is the only available raw mouse usage data set. 
We are actively working on a collection protocol and system in order to be able to create our own mouse dynamics data set for intruder detection. 

\bibliographystyle{unsrt}
\bibliography{refs}

\end{document}